\documentclass[twocolumn,english,aps,prl,reprint, superscriptaddress,showpacs,longbibliography,showkeys]{revtex4-2}
\usepackage{amsmath,amssymb,bbm,mathrsfs,bm,braket,color,graphicx,comment,xcolor,dsfont,multirow}
\usepackage[colorlinks,citecolor=blue,urlcolor=blue,linkcolor = blue]{hyperref}
\usepackage[mathscr]{euscript}
\usepackage{subfigure}
\usepackage{multirow}
\usepackage[shortlabels]{enumitem}

\usepackage{tikz, ifthen}
\usetikzlibrary{arrows.meta}
\usetikzlibrary{decorations.pathreplacing,calc}
\usepackage{float}
\makeatletter
\let\newfloat\newfloat@ltx
\makeatother
\usepackage{algorithm}
\usepackage[noend]{algpseudocode}

\begin{document}
\title{Universal Parity Quantum Computing}
\author{Michael Fellner}
\affiliation{Institute for Theoretical Physics, University of Innsbruck, A-6020 Innsbruck, Austria}
\affiliation{Parity Quantum Computing GmbH, A-6020 Innsbruck, Austria}
\author{Anette Messinger}
\affiliation{Parity Quantum Computing GmbH, A-6020 Innsbruck, Austria}
\author{Kilian Ender}
\affiliation{Institute for Theoretical Physics, University of Innsbruck, A-6020 Innsbruck, Austria}
\affiliation{Parity Quantum Computing GmbH, A-6020 Innsbruck, Austria}
\author{Wolfgang Lechner}
\affiliation{Institute for Theoretical Physics, University of Innsbruck, A-6020 Innsbruck, Austria}
\affiliation{Parity Quantum Computing GmbH, A-6020 Innsbruck, Austria}
\date{\today}

\begin{abstract}
We propose a universal gate set for quantum computing with all-to-all connectivity and intrinsic robustness to bit-flip errors based on the parity encoding. We show that logical controlled phase gates and $R_z$ rotations can be implemented in the parity encoding with single-qubit operations. Together with logical $R_x$ rotations, implemented via nearest-neighbor controlled-NOT gates and an $R_x$ rotation, these form a universal gate set. As the controlled phase gate requires only single-qubit rotations, the proposed scheme has advantages for several cornerstone quantum algorithms, e.g.\ the quantum Fourier transform. We present a method to switch between different encoding variants via partial on-the-fly encoding and decoding.
\end{abstract}

\maketitle


Designing quantum computers \cite{Feynman1982,Reck1994, cirac1995quantum,DiVincenzo1995,  Barenco1995_Elementary, Lloyd1996,  Bennett1997, Knill1998, jaksch2000fast,  Raussendorf2001, Knill2001, Nakamura1999, Molmer1999, kitaev2002, Koch2007, Haffner2008, Terhal2015} and quantum algorithms \cite{Deutsch1992, Berthiaume1994, Grover1996, Shor1997, Bernstein1997, Mosca1999, Brassard2002, vanDam2002, Harrow2009} is a current grand challenge in science and engineering, motivated by the prospect of solving certain problems exponentially faster than any known classical algorithms \cite{Shor1997}. However, the fundamental rules of quantum mechanics that make this new paradigm possible also impose fundamental restrictions. 
In contrast to classical information, quantum information cannot be copied, which is known as the no-cloning theorem, but only propagated \cite{Lieb1972}.
Thus, quantum computers will not be able to follow the von Neumann architecture \cite{vonNeumann1993} with separated memory and computational unit. As the quantum CPU serves as memory and computational unit at the same time, connectivity between any quantum bits on the chip is required. In current standard approaches to gate-based quantum computers, either these long-range interactions are implemented as physical interactions, which limits scalability, or quantum information is moved on the chip via SWAP sequences, which requires a large overhead in gates . Although there are recent approaches toward qubit routing that address this issue \cite{Bapat2022, Devulapalli2022}, exchanging information between qubits remains a challenging problem. 

In this Letter, we propose a novel universal quantum computing approach based on the Lechner-Hauke-Zoller (LHZ) architecture \cite{Lechner2015}, which was originally designed for quantum annealing. In this parity-based paradigm, each physical qubit represents the parity of multiple logical qubits. We extend the LHZ architecture, up to now only used for solving combinatorial optimization problems, to a universal quantum computing approach by providing a universal gate set on parity-encoded states and with that, open up new possibilities for universal quantum computation.
These extensions include an additional row of \textit{data qubits} added to the original LHZ layout to enable control of single logical qubits. We introduce logical operations, in particular $R_x$ rotations, to establish a universal gate set in the logical space. 
As the parity constraints no longer need to be enforced throughout the computation, they can be utilized for error correction.

\begin{figure*}
    \centering
    \scriptsize
        \begin{tabular}{c||c|c||c|c}
        \multirow{2}{*}{Gate} & \multirow{2}{*}{Standard Circuit} & Schematic Standard  & \multirow{2}{*}{Circuit in LHZ Scheme} & Schematic Implementation \\
        & & Implementation & & in LHZ\\\hline
       $R_x^{(i)}(\varphi)$ & \parbox[m]{4.3cm}{\includegraphics[scale=.65]{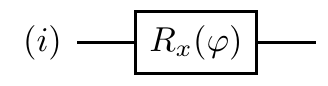}} &
       \parbox[m]{3.5cm}{\vspace{1.5mm}\includegraphics{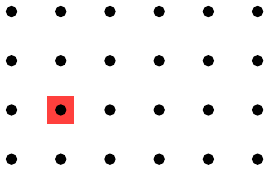}\vspace{1.5mm}} &
       \parbox[m]{4.3cm}{\includegraphics[scale=.65]{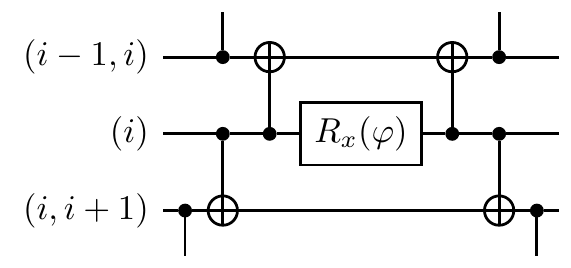}} & 
       \parbox[m]{3.8cm}{\includegraphics{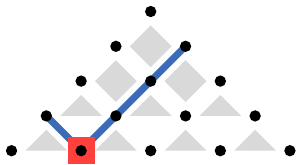}} 
       \\ \hline
       $R_z^{(i)}(\theta)$  & \parbox[m]{4.3cm}{\includegraphics[scale=.65]{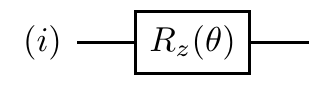}} &  
       \parbox[m]{3.5cm}{\vspace{1.5mm}\includegraphics{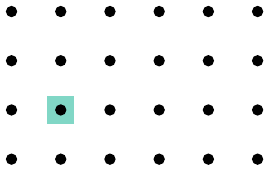}\vspace{1.5mm}} &
       \parbox[m]{4.3cm}{\includegraphics[scale=.65]{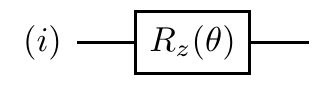}}   &
       \parbox[m]{3.8cm}{\includegraphics{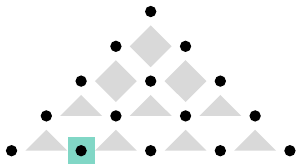}} \\ \hline
       $\text{CP}^{(i,j)}_\phi$ & \parbox[m]{4.3cm}{\includegraphics[scale=.65]{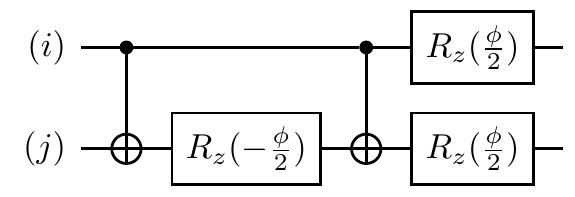}} &
       \parbox[m]{3.5cm}{\vspace{1.5mm}\includegraphics{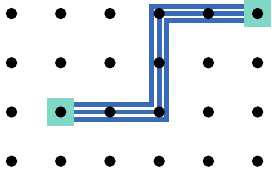}} &
       \parbox[m]{4.3cm}{\includegraphics[scale=.65]{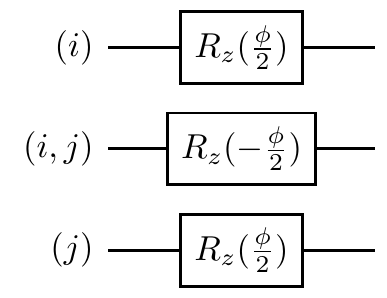}}  &
       \parbox[m]{3.8cm}{\includegraphics{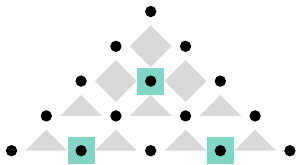}} 
    \end{tabular}
    
    \caption{
    Overview over the implementation of a universal gate set in the LHZ scheme. All operations have been decomposed to CNOT gates and local rotations.
    Blue lines represent a chain of CNOT gates, while red (darker) and green (lighter) squares depict local $ R_x$ and $ R_z$ rotations, respectively. Triple lines correspond to SWAP gates, consisting of 3 CNOT gates each.}
    \label{fig:basis_gates}
\end{figure*}

As it only requires nearest-neighbor interactions between qubits on a square lattice chip, our proposal can be implemented on state-of-the-art quantum devices, independent of the qubit platform. Suitable platforms are for example superconducting qubits~\cite{wallraff2004strong, Koch2007, Houck2008, Barends2013, arute2019quantum, Wu2021}, neutral atoms \cite{saffman2010quantum, henriet2020quantum, rydberg_cong_2022}, or trapped ions~\cite{blatt2012quantum, kielpinski2002architecture, Lekitsch2017, ion_hughes_2020}.
We show that the parity transformation renders diagonal multiqubit operators between arbitrary logical qubits into single-qubit physical gates and, in turn, nondiagonal logical operators into sequences of physical gates. The transformation eliminates the need for long-range interactions and thus SWAP gates as illustrated in Fig.~\ref{fig:basis_gates}. Furthermore, redundant encoding offers the potential for intrinsic tolerance against bit-flip errors. 
We additionally present a possibility to choose and switch between different variants of the parity mapping, containing subsets of parity qubits tailored to the algorithmic requirements. This allows for further reduction of computational resources. 

The gate set presented here contains operations that correspond to $R_x$  and $R_z$ rotations and controlled phase gates acting on logical qubits. Because of the absence of any connectivity limitations, we expect this approach to have an impact on the design of next generation quantum devices \cite{Preskill_2018}. The scheme allows for an efficient implementation of controlled rotations around the $z$ axis and is thus advantageous for the quantum Fourier transform \cite{NielsenChuang2011} which is the basis of Shor's factoring algorithm \cite{Shor1997} as well as quantum addition \cite{Draper2000}. Implementations of well-known quantum algorithms in the parity architecture are shown in detail in the associated publication Ref.~\cite{applications_paper}.\\

\begin{figure}
    \centering
    \includegraphics{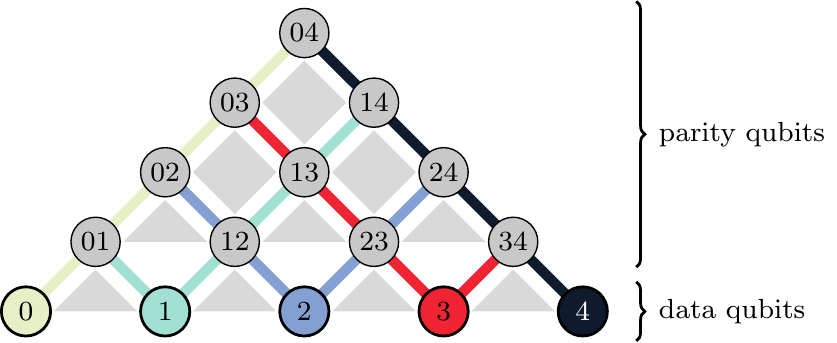}
    \caption{Illustration of the modified LHZ architecture with logical lines. Three- and four-body constraints are represented by light gray triangles and squares between corresponding qubits. Data qubits with single logical indices are added as an additional row at the bottom of the architecture to allow direct access to logical $R_z$ rotations. Colored lines connect all qubits whose labels contain the same logical index. Logical $R_x$ rotations can be realized with chains of CNOT gates along the corresponding line.}
    \label{fig:architecture}
\end{figure}

\paragraph{Parity mapping}
The LHZ architecture \cite{Lechner2015} expands the Hilbert space of $n$ logical qubits to a Hilbert space of ${K=n(n-1)/2}$ physical qubits (parity qubits, with operators ${\sigma}$), encoding the parity of pairs of logical qubits (with operators $\tilde\sigma$) such that for any state $\ket{\psi}$ in the code space,
\begin{equation}\label{eq:parity}
    \tilde\sigma_{z}^{(i)}\tilde\sigma_{z}^{(j)}\ket{\psi} =  \sigma_z^{(ij)}\ket{\psi},
\end{equation}
where the superscripts correspond to qubit labels. The code space is restricted to configurations corresponding to valid logical states $\ket{\psi}$ by ${K-n+1}$ parity constraints of the form
\begin{equation}
    C_l \ket{\psi} :=  \sigma_z^{(l_1)}\sigma_z^{(l_2)}\sigma_z^{(l_3)}\,[\sigma_z^{(l_4)}]\ket{\psi} = \ket{\psi}.
\end{equation}
where the indices $l_i$ correspond to pairs of logical indices, such that in every constraint, each logical index appears an even number of times. The brackets around $\sigma_z^{(l_4)}$ indicate that a constraint can contain either 3 or 4 qubits. Here, we use a slightly modified layout compared with the original LHZ layout, shown in Fig.~\ref{fig:architecture}, with an additional row of physical qubits which have a direct correspondence to single logical qubits,
\begin{equation}
    \tilde\sigma_{z}^{(i)} =  \sigma_z^{(i)}.
\end{equation}
In the following, these additional qubits are referred to as \textit{data qubits}. As depicted in Fig.~\ref{fig:architecture}, $n$ all-to-all connected logical qubits are represented by $K=n(n+1)/2$ physical qubits and $K-n$ constraints.
The parity constraints generate the stabilizer of the code space \cite{Gottesman1997}, additionally allowing for detection of bit-flip errors and thus an intrinsic fault tolerance of the encoding.\\

\paragraph{Logical qubits and operators}\label{sec:logical_lines}
Next, we introduce the concept of \textit{logical lines} denoted as $Q_i$, which have been identified in a different context in Ref.~\cite{Rocchetto2016_stabilizer}. A logical line $Q_i$ is defined as the set of all parity qubits containing the logical index $i$. In the LHZ architecture, qubits that contain a particular index are arranged along lines, which are indicated as solid lines in Fig.~\ref{fig:architecture} to guide the eye.
The red line for example extends from the data qubit $(3)$ to all parity qubits that contain the index $3$, and thus contain all relative parity information with respect to the logical qubit $(3)$.
For each logical line $Q_i$ we define an operator
\begin{equation}
    \tilde\sigma_x^{(i)}={\sigma}_x^{(i)} \prod_{j<i}{\sigma}_x^{(ji)}\prod_{j>i}{\sigma}_x^{(ij)}
\end{equation}
acting on every qubit in the respective line.
The operators $\tilde\sigma_x^{(i)}$ and $\tilde\sigma_z^{(i)}$ commute with all parity constraints and fulfill the (anti-) commutation relations for Pauli operators
\begin{equation}
    \{\tilde\sigma_x^{(i)}, \tilde\sigma_z^{(i)}\}=[\tilde\sigma_x^{(i)}, \tilde\sigma_z^{(j)}]=0
\end{equation}
for ${i\neq j}$. We can therefore identify the operators with Pauli operators acting on logical qubits, and build arbitrary logical rotations as
\begin{equation}\label{eq:rx_implementation}
  \tilde R_x(\alpha) =\exp\left(-i\frac{\alpha}{2} \tilde\sigma_x^{(i)}\right)
\end{equation}
and
\begin{equation}\label{eq:rz_implementation}
    \tilde R_z(\alpha) =\exp\left(-i \frac{\alpha}{2} \tilde\sigma_z^{(i)}\right).
\end{equation} 
Throughout this Letter, tildes are used to denote operators with the corresponding action on logical qubits.
While a logical $\tilde R_z$ rotation can be directly implemented using the respective physical operator on the data qubit, the $\tilde R_x$ operator on logical qubit $(i)$ given by expression \eqref{eq:rx_implementation} can be implemented via physical controlled-NOT (CNOT) gates along the logical line $Q_i$ and a physical single-body $R_x$ rotation, as depicted in Fig.~\ref{fig:basis_gates} (also see Ref.~\cite{Cowtan2020} for details on efficient implementations of exponentials of multiqubit Pauli operators). The chosen layout further ensures that all operators associated with the logical lines can be implemented with local operations and nearest-neighbor CNOT gates on a square lattice.\\

\paragraph{Encoding and decoding}
In the following we discuss the encoding and decoding of arbitrary states from and to the data qubits. Let us first consider the trivial logical state $\ket{0}^{\otimes n}$ which can be encoded in the LHZ scheme  by preparing all physical qubits in the product state $\ket{0}^{\otimes K}$. This state fulfills all constraints, and is the joint eigenstate of the logical $\tilde \sigma_z$ operators with eigenvalue $+1$. Encoding an arbitrary, unknown quantum state into the LHZ scheme is less straightforward. 
Classically, the states of the data qubits are identical to the corresponding logical qubits (by definition, a measurement of these qubits in the $z$ basis corresponds to a measurement of the logical state). However, they are typically highly entangled with the other qubits, and simply tracing out the parity qubits would cause a loss of coherence and therefore phase information.

\begin{figure}
    \centering
    \includegraphics{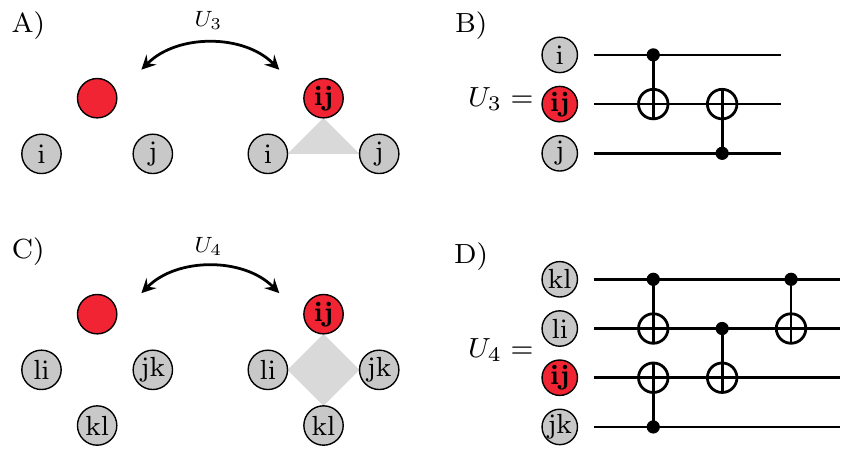}
    \caption{Encoding and decoding circuits to add or remove a qubit and the corresponding constraint to or from the code. A) Qubit $(ij)$ is directly encoded via the adjacent data qubits $(i)$ and $(j)$ in a three-body constraint, with the circuit B). C) Qubit $(ij)$ is encoded using other parity qubits in a four-body constraint, with the circuit D). A qubit without a label indicates a qubit without any parity information, being in the state $\ket{0}$.
    }
    \label{fig:encoding_circuit}
\end{figure}

We thus introduce an encoding and decoding strategy to add or remove an arbitrary number of parity qubits to or from the code.
Suppose we start with the logical qubits, each of them encoded in a data qubit. We can now add a physical qubit $(ij)$ corresponding to the parity of qubits $(i)$ and $(j)$, by initializing this qubit in the state $\ket{0}$ and then imposing the parity on it with two CNOT gates controlled by qubits $(i)$ and $(j)$, as shown in Fig.~\ref{fig:encoding_circuit}(B).
This procedure adds an additional parity qubit to the code and guarantees that the corresponding constraint ${C=\sigma_z^{(i)}\sigma_z^{(j)}\sigma_z^{(ij)}}$ is satisfied.

Following this procedure, it is possible to add parity qubits and constraints by applying this strategy iteratively. Instead of directly obtaining the parity information from the data qubits, one can also use the parity qubits included in local constraints, as for example the plaquettes shown in Fig.~\ref{fig:architecture}: By definition, the parity of all but one qubit of a constraint is always encoded in the state of the remaining qubit, i.e.\
\begin{align*}
    \sigma_z^{(ij)}\sigma_z^{(jk)}\sigma_z^{(kl)}\sigma_z^{(li)}\ket\psi = \ket\psi\\
    \Rightarrow \sigma_z^{(ij)}\ket\psi = \sigma_z^{(jk)}\sigma_z^{(kl)}\sigma_z^{(li)}\ket\psi
\end{align*}
This transition and the encoding circuits are shown in Figs.~\ref{fig:encoding_circuit}(C)-\ref{fig:encoding_circuit}(D).

Conversely, applying the same gate sequence to a parity-encoded set of qubits removes the targeted qubit from the code and projects it to the state of the corresponding constraint. If the constraint was fulfilled, this is the state $\ket{0}$. For a layout as depicted in Fig.~\ref{fig:architecture} with $n$ logical qubits, this procedure allows encoding and decoding circuits with an overall circuit depth of $n+1$. The exact circuit is given in the Supplemental Material. 

The described procedure does not only offer a way to build up or collapse the all-to-all connected LHZ scheme, but also holds the possibility to switch between different variants of the parity mapping and adapt the number of parity qubits to the algorithmic requirements during computation.
For example, if some interactions are not needed for a certain algorithm, it is not always necessary to have the corresponding parity qubits in the code. A reduction in the number of parity qubits results in shorter logical lines and thus fewer physical gates.
In addition to the two-body parities as introduced in Eq.~\eqref{eq:parity}, it is also possible to encode higher-order $k$-body parity qubits in the same manner, when using suitable layouts (see Ref.~\cite{applications_paper} for more details). As an example, a three-body parity qubit can be used to enable a nontrivial interaction between three logical qubits. \\

\paragraph{Universal gate set}
Single-qubit operations on logical qubits can be constructed from the operators introduced in Eqs.~\eqref{eq:rx_implementation}-\eqref{eq:rz_implementation} using the decomposition
\begin{equation}\label{eq:unitary_decomposition}
    U= R_{z}(\alpha)R_{x}(\beta)R_{z}(\gamma).
\end{equation}
To obtain a universal gate set, an additional two-qubit entangling gate is necessary.
In the LHZ encoding, a native logical two-qubit operation is obtained by performing a single $ R_z$ rotation on a physical parity qubit,
\begin{equation}
    {R}_z^{(ij)}(\alpha)=\exp\left(-i\frac{\alpha}{2} {\sigma}_z^{(ij)}\right),
\end{equation}
which is stabilizer equivalent (i.e.,\ the same up to the application of a constraint operator ${C=\sigma_z^{(i)}\sigma_z^{(j)}\sigma_z^{(ij)}}$) to the operator 
\begin{equation}
    \exp\left( -i\frac{\alpha}{2} \sigma_z^{(i)}\sigma_z^{(j)} \right)
\end{equation}
and thus effectively performs the two-body operation
$
    \exp\left(-i\frac{\alpha}{2} \tilde\sigma_{z}^{(i)}\tilde\sigma_{z}^{(j)}\right)
$
on the logical qubits, as obvious from Eq.~\eqref{eq:parity}.
This operation can be transformed to a logical controlled phase gate $\text{CP}_\phi$ with local $ R_z$ rotations only, as shown in the following. For the sake of simplicity, qubit indices are omitted.
We start from an operation
\begin{equation}
    e^{-i\frac{\alpha}{2} \sigma_z\otimes\sigma_z} = 
    \text{diag}\left(e^{-i\frac{\alpha}{2}}, e^{i\frac{\alpha}{2}}, e^{i\frac{\alpha}{2}} , e^{-i\frac{\alpha}{2}}\right)
\end{equation}
and the single-body $R_z$ rotation,
\begin{equation}
    R_z(\theta) = e^{-i\frac{\theta}{2} \sigma_z} = \text{diag}(
        e^{-i\frac{\theta}{2}}, e^{i\frac{\theta}{2}}),
\end{equation}
and define
\begin{equation}\label{eq:u_r}
    U_R := [\mathds{1}\otimes R_z(\beta)]e^{-i\frac{\alpha}{2} \sigma_z\otimes\sigma_z}[R_z(\gamma)\otimes\mathds{1}].
\end{equation}
Evaluating  Eq.~\eqref{eq:u_r} for ${-\alpha=\beta=\gamma= \frac{\phi}{2}}$ yields
\begin{equation*}
    U_R = e^{-i\frac{\phi}{4}} 
    \cdot \text{diag}(1, 1, 1, e^{i\phi})= e^{-i\frac{\phi}{4}}\cdot  \text{CP}_{\phi},
\end{equation*}
which corresponds, up to a global phase, to the controlled phase gate $\text{CP}_{\phi}$. Using the identities defined above, this can be implemented in our scheme with local operations on the physical parity qubits and data qubits as
\begin{equation}
    \text{C} \tilde{\text{P}}^{(i, j)}_{\phi} = {R}_z^{(i)}\left(\frac{\phi}{2}\right){R}_z^{(ij)}\left(-\frac{\phi}{2}\right){R}_z^{(j)}\left(\frac{\phi}{2}\right).
\end{equation}
Here, $i$ and $j$ are the indices of the involved logical qubits and ${R}_z$ indicates the rotation on the corresponding data or parity qubit. In particular, for ${\phi=\pi}$, we obtain the CZ gate (controlled Z gate). This can in turn be transformed to a CNOT gate, by applying logical Hadamard gates before and after the CZ gate on the desired target qubit.
The operations $\tilde R_x$, $\tilde R_z$ and $\text{C}\tilde{\text P}_{\phi}$ form a \textit{universal gate set} in the LHZ scheme and we can not only build arbitrary quantum circuits by applying CNOT gates and controlling local fields (parameters only occur in single-qubit operations), but also exploit the comparably simple implementation of a controlled phase gate. The standard gate model requires two CNOT- and three single-qubit gates to implement a controlled phase gate, as depicted in Fig.~\ref{fig:basis_gates}. Platforms with limited connectivity typically need a large number of SWAP gates in addition.
In contrast, a logical controlled phase gate in our approach only requires three parallel single-body rotations.
The required resources for implementing common gates and gate sequences in our scheme are listed in Table~\ref{tab:required_resources}. The depth and the gate count for nondiagonal single-body operations result from the CNOT chains involved.
The main advantage of the encoding clearly comes from the depth-1 implementation of diagonal gates.
Note that while nondiagonal operations have a higher cost, their increased weight allows for intrinsic tolerance to bit-flip errors as an additional advantage.
\begin{table}
\begin{tabular}{c|c|c|c}
 \multirow{2}{*}{Logic gate} & \multicolumn{3}{c}{Required gates in LHZ}\tabularnewline 
 & Single qubit & 2 qubit & Depth \tabularnewline \hline
$R_x$ & $1$ & $2(n-1)$ & $2\left\lceil \frac{n}{2}\right\rceil+1$ \tabularnewline 
$\sigma_x$ & $n$ & $0$ & $1$ \tabularnewline 
$R_z$ & $1$ & $0$ & $1$ \tabularnewline 
$U$& $3$ & $2(n-1)$ & $2\left\lceil \frac{n}{2}\right\rceil+1$\footnote{For ${n\leq 4}$, the depth increases by two steps because the $R_z$ rotations cannot be performed parallel to other operations.}\tabularnewline
CP & $3$ & $0$ & $1$\tabularnewline 
$\text{CNOT}^{(c,t)}$ & $7$ & $2(n-1+|c-t|)$ & $\leq 4\left\lceil\frac{n}{2}\right\rceil + 3$\footnote{Can be further optimized depending on the control qubit and target qubit. For a detailed discussion, see Ref.~\cite{applications_paper}.}\tabularnewline
$\prod_{i=1}^m U_i^{(n_i)}$\footnote{$n_i\leq n$ denotes the qubit on which $U_i$ acts and ${m\leq n}$ is the number of qubits involved. We count consecutive single-qubit operations as a single time step.}& $3m$ & $2n(n-1)$ & $ 2n+3$\tabularnewline
\end{tabular}
\caption{The number of physical operations needed to perform logical gates in the parity encoding for $n>4$. All operations were decomposed into CNOT gates and local $R_x$ and $R_z$ rotations. The spin-flip operator $\sigma_x$ is a special case of the $R_x$ rotation, for which the logical implementation reduces to a product of single-qubit spin flips which can be performed in parallel. The unitary $U$ represents an arbitrary nondiagonal single-qubit operation. A detailed derivation of the given numbers can be found in Ref.~\cite{applications_paper}.}
\label{tab:required_resources}
\end{table}

Following Eq.~\eqref{eq:unitary_decomposition}, any local unitary operation $U$ on a logical qubit $(i)$ can be performed in the LHZ scheme by performing it on the data qubit $(i)$ and surrounding it by the CNOT chains as in the logical $\tilde R_x$ rotation,
\begin{align*}
    \tilde U
    &= \underbrace{R_z^{(i)}(\alpha)}_{\tilde R_z}\,\underbrace{...\text{CNOT}^{(ij)} R_x^{(i)}(\beta) \text{CNOT}^{(ij)}...}_{\tilde R_x}\,\underbrace{R_z^{(i)}(\gamma)}_{\tilde R_z} \\
    &= ...\text{CNOT}^{(ij)} R_z^{(i)}(\alpha) R_x^{(i)}(\beta) R_z^{(i)}(\gamma) \text{CNOT}^{(ij)}... \\
    &= ...\text{CNOT}^{(ij)} U^{(i)} \text{CNOT}^{(ij)}...
\end{align*}
The circuit depth for this construction is given in Table~\ref{tab:required_resources}.
Fixing the qubit on which the $R_x$ rotations are performed to the respective data qubit reduces hardware requirements such that $R_x$ rotations are only ever necessary on the data qubits, while for the parity qubits, local $R_z$ rotations and CNOT gates between them are sufficient.

Note that a circuit implementing a product of ${m\leq n}$ logical single-qubit operators can be realized by applying the decoding circuit, performing the single-body operations on the data qubits and encoding again with an overall circuit depth of ${2n+3}$ and a CNOT-gate count of ${2n(n-1)}$. \\

\paragraph{Error correction}
The redundancy of the parity encoding allows for correction or mitigation of bit-flip errors. Because all parity constraints commute with any logical operator, measuring their value does not disturb the logical state of the system. Constraint measurements can be performed either with the help of ancillary measurement qubits in the center of each plaquette \cite{NielsenChuang2011, Fowler2012}, or by applying the decoding circuit as in Figs.~\ref{fig:encoding_circuit}(B) and \ref{fig:encoding_circuit}(D) to each constraint, measuring the decoded qubit, and encoding it again.

As the transition from one logical basis state to another flips $n$ physical qubits, it is in principle possible to correct for multiple errors at a time. For different encoding variants, for example with a reduced number of parity qubits, the number of simultaneously correctable errors depends on the number of qubits in the shortest logical line. F.~Pastawski and J.~Preskill successfully demonstrated error correction via belief propagation on a LHZ chip \cite{Pastawski2016}, and showed that the LHZ encoding is robust against weakly correlated bit-flip noise. An analysis of the bit-flip error-correction capability of the LHZ architecture is provided in the Supplemental Material.

To ideally complement the bit-flip tolerance of the LHZ encoding, it is advisable to use physical qubits which are intrinsically robust against phase errors \cite{Aliferis2009, Puri2020, Lescanne2020, Lee2021}.\\

\paragraph{Conclusion and outlook}\label{sec:conclusion}
In conclusion, we have demonstrated a universal gate set implemented in the LHZ encoding, opening up a new strategy for universal quantum computation. All gates can be implemented on state-of-the-art quantum devices that fulfill the comparably low requirement of nearest-neighbor connectivity on a square lattice. Furthermore, the encoding can be dynamically adjusted by adding or removing parity qubits tailored to the requirements of different algorithms. The introduced decoding scheme can be used to decode the output of optimization algorithms run in the parity scheme as for example proposed in Refs.~\cite{lechner2020, ender2022modular, Sieberer2018}, in order to further work with the resulting quantum states. With its availability of nonlocal and multiqubit operations and the intrinsic error-correction capability, we expect our work to be a step towards the next generation of quantum computers.

\paragraph{Acknowledgements -} Work at the University of Innsbruck is supported by the Austrian Science Fund (FWF) through a START grant under Project No. Y1067-N27 and the Special Research Programme (SFB) BeyondC Project No. F7108-N38. This work was supported by the Austrian Research Promotion Agency under Grant (FFG Project No. 892576, Basisprogramm).

%

\end{document}


\title{Supplemental material for ``Universal Parity Quantum Computing''}
\author{Michael Fellner}
\affiliation{Institute for Theoretical Physics, University of Innsbruck, A-6020 Innsbruck, Austria}
\affiliation{Parity Quantum Computing GmbH, A-6020 Innsbruck, Austria}
\author{Anette Messinger}
\affiliation{Parity Quantum Computing GmbH, A-6020 Innsbruck, Austria}
\author{Kilian Ender}
\affiliation{Institute for Theoretical Physics, University of Innsbruck, A-6020 Innsbruck, Austria}
\affiliation{Parity Quantum Computing GmbH, A-6020 Innsbruck, Austria}
\author{Wolfgang Lechner}
\affiliation{Institute for Theoretical Physics, University of Innsbruck, A-6020 Innsbruck, Austria}
\affiliation{Parity Quantum Computing GmbH, A-6020 Innsbruck, Austria}
\date{\today}

\begin{abstract}
    In this supplemental material we present details on the circuit for the encoding and decoding of states for universal parity quantum computing with a depth of ${n+1}$. Furthermore, we provide details on the robustness against bit-flip errors and respective correction capabilities of the encoding. 
\end{abstract}

\maketitle
\section{Low-depth circuit for encoding and decoding}
\begin{figure}
    \centering
\includegraphics{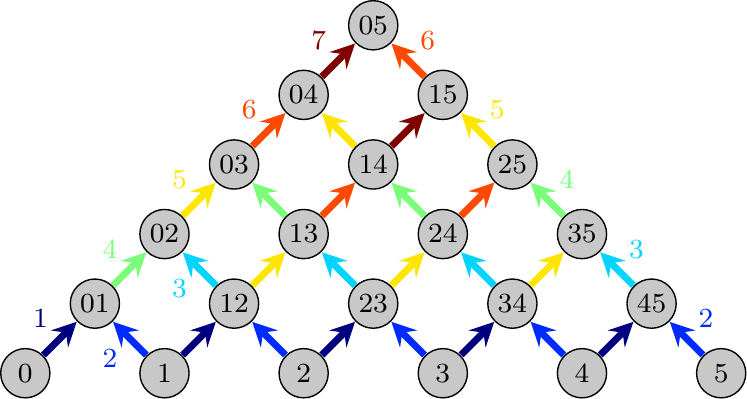}
    \caption{Encoding gate sequence for an LHZ architecture with ${n=6}$ logical qubits, using ${n+1}$ time steps. Arrows represent CNOT gates pointing at the target qubit. Colors (from dark blue to dark red) indicate the temporal ordering of the gates.}
    \label{fig:encoding}
\end{figure}

A routine for encoding a state in the LHZ scheme starting from $n$ physical data qubits (corresponding to the logical qubits) is obtained from iteratively applying the circuits to encode an additional parity qubit, as discussed in the main text. Commutation and cancellation of CNOT gates reduces the circuit depth to ${n+1}$. This protocol with depth ${n+1}$ for encoding is given by Algorithm~\ref{alg:encoding}, where data qubits and parity qubits are denoted $(i)$ and $(i, j)$, respectively, and visualized in Fig.~\ref{fig:encoding}. Arrows representing CNOT gates point from the control to the target qubit. All gates within a single step can be executed in parallel. Note that the reversed sequence can be used to decode all information onto the data qubits.

\begin{algorithm}
\caption{Encoding sequence with depth $n+1$}\label{alg:encoding}
\begin{algorithmic}
\For{$0 \leq i < n-1$}\tikzmark{top1} 
\State $(i) \overset{\rm CNOT}{\longrightarrow} (i, i+1)$
\EndFor\tikzmark{bottom1}

\For{$0 \leq i < n-1$}\tikzmark{top2}
\State $(i+1) \overset{\rm CNOT}{\longrightarrow} (i, i+1)$
\EndFor\tikzmark{bottom2}

\For{$1\leq i < n-1$}\tikzmark{top3}
\State $(i, i+1) \overset{\rm CNOT}{\longrightarrow} (i-1, i+1)$
\EndFor\tikzmark{bottom3}

\State $(0, 1) \overset{\rm CNOT}{\longrightarrow} (0, 2)$\tikzmark{top4}
\For{$1\leq i < n-2$}
\State $(i, i+2) \overset{\rm CNOT}{\longrightarrow} (i-1, i+2)$
\EndFor\tikzmark{bottom4}

\For{$3\leq j<n$}
    \State $(0,j-1) \overset{\rm CNOT}{\longrightarrow} (0, j)$ \tikzmark{top5}
    \State $(1,j-1) \overset{\rm CNOT}{\longrightarrow} (1, j)$
    \For{$1\leq i < n-j$}
        \State $(i+1, i+j-1) \overset{\rm CNOT}{\longrightarrow} (i+1, i+j)$\hspace{2mm}\tikzmark{right}
        \State $(i, i+j) \overset{\rm CNOT}{\longrightarrow} (i-1, i+j)$
    \EndFor\tikzmark{bottom5}

\EndFor
\end{algorithmic}
\AddNote{top1}{bottom1}{right}{Step $1$}
\AddNote{top2}{bottom2}{right}{Step $2$}
\AddNote{top3}{bottom3}{right}{Step $3$}
\AddNote{top4}{bottom4}{right}{Step $4$}
\AddNote{top5}{bottom5}{right}{Step $j+2$}
\end{algorithm}

\section{Suppression of bit-flip errors}
In order to examine the capability of the encoding for correcting bit-flip errors, we conduct an analysis of the probability of non-correctable errors based on statistical arguments. In general, the code distance $d$ is given by the number of qubits in the shortest logical line, and up to $(d-1)/2$ simultaneous bit-flips can be corrected. If more details about the geometry and physical error rates are known, it is in principle possible to also correct higher numbers of errors on some occasions but we do not assume this in the following.
For our analysis we consider the extended LHZ layout with $n$ data qubits and ${n(n-1)/2}$ parity qubits, yielding a code distance of $d=n$. We measure each parity constraint $n$ times using a separate measurement ancilla to suppress time-like logical errors as well. In order to estimate the error-robustness, we calculate the probability of a logical error caused by at least one of the following scenarios occurring during a cycle of $n$ syndrome measurements:
\begin{enumerate}[(a)]
    \item more than ${(n-1)/2}$ bit-flips occur between two syndrome measurements or
     \item more than half of the syndrome measurement repetitions on a constraint yield a faulty result.
\end{enumerate}
For the sake of simplicity we assume every constraint to contain 4 qubits and every qubit to be part of 4 constraints. This is the case in the limit of large $n$ and otherwise leads to a valid upper bound for the logical error rates.

\begin{figure*}
    \centering
    \includegraphics[width=.95\textwidth]{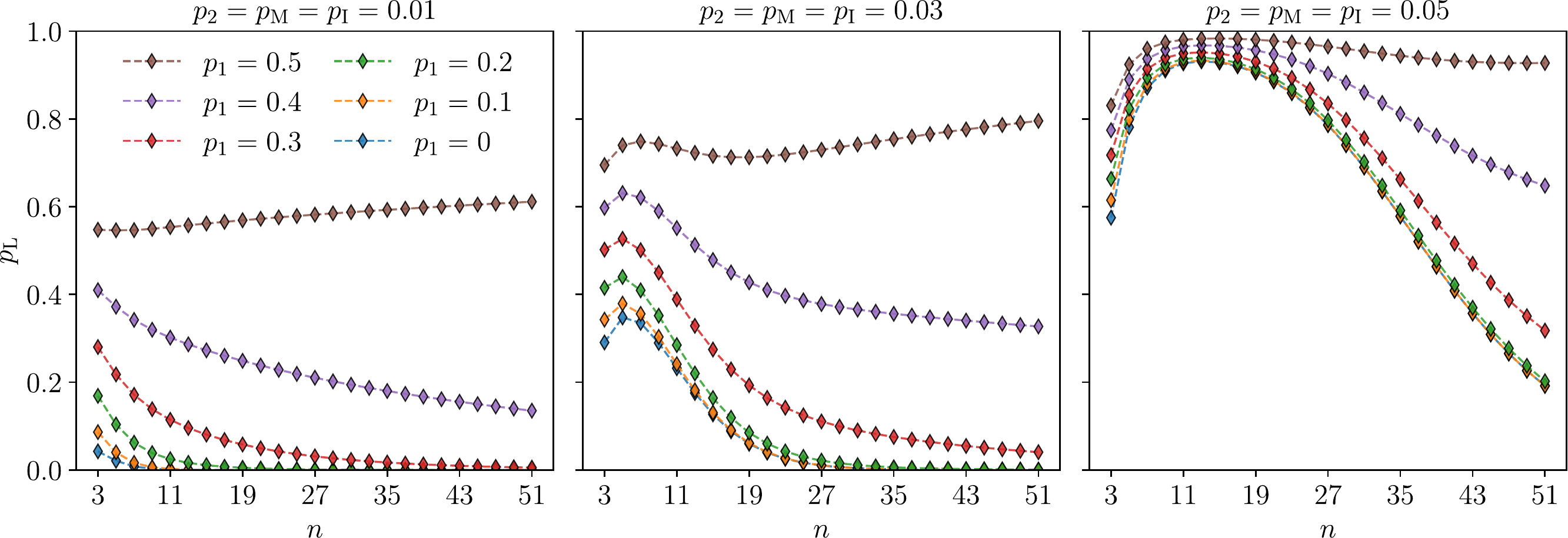}
    \caption{Total probability for logical bit-flip errors to occur after error correction for different physical error rates. For sufficiently small single-qubit error rates, the logical error rate decreases exponentially with $n$ after an initial increase.}

    \label{fig:error_simulations}
\end{figure*}

A syndrome measurement then consists of 4 CNOT gates from the parity qubits to the ancilla qubit, and one initialization and one measurement of the ancilla qubit. 
We assume every physical qubit to be subject to a bit-flip error (regardless of its source) with probability $p_1$ before the syndrome measurement. 
We further assume that each qubit involved in a CNOT gate is subject to a bit-flip error with probability $p_2$. The probabilities for faulty qubit initialization and measurement are denoted by $p_{\rm I}$ and $p_{\rm M}$, respectively. Note that these error-rates are only relevant in the syndrome measurement process, all other error sources are covered by $p_1$. In our analysis, we set ${p_2=p_\text{I}=p_\text{M}}$ because they are usually on the same scale for most practical implementations. Fig.~\ref{fig:error_simulations} shows the probability for the occurrence of at least one logical error, as described above, after a sequence of syndrome measurements for various combinations of physical error probabilities. The results of our calculations show a qualitative agreement with the performance of the belief propagation in the LHZ scheme presented in Ref.~\cite{Pastawski2016}.
\begin{figure}
    \centering
    \includegraphics[width=\columnwidth]{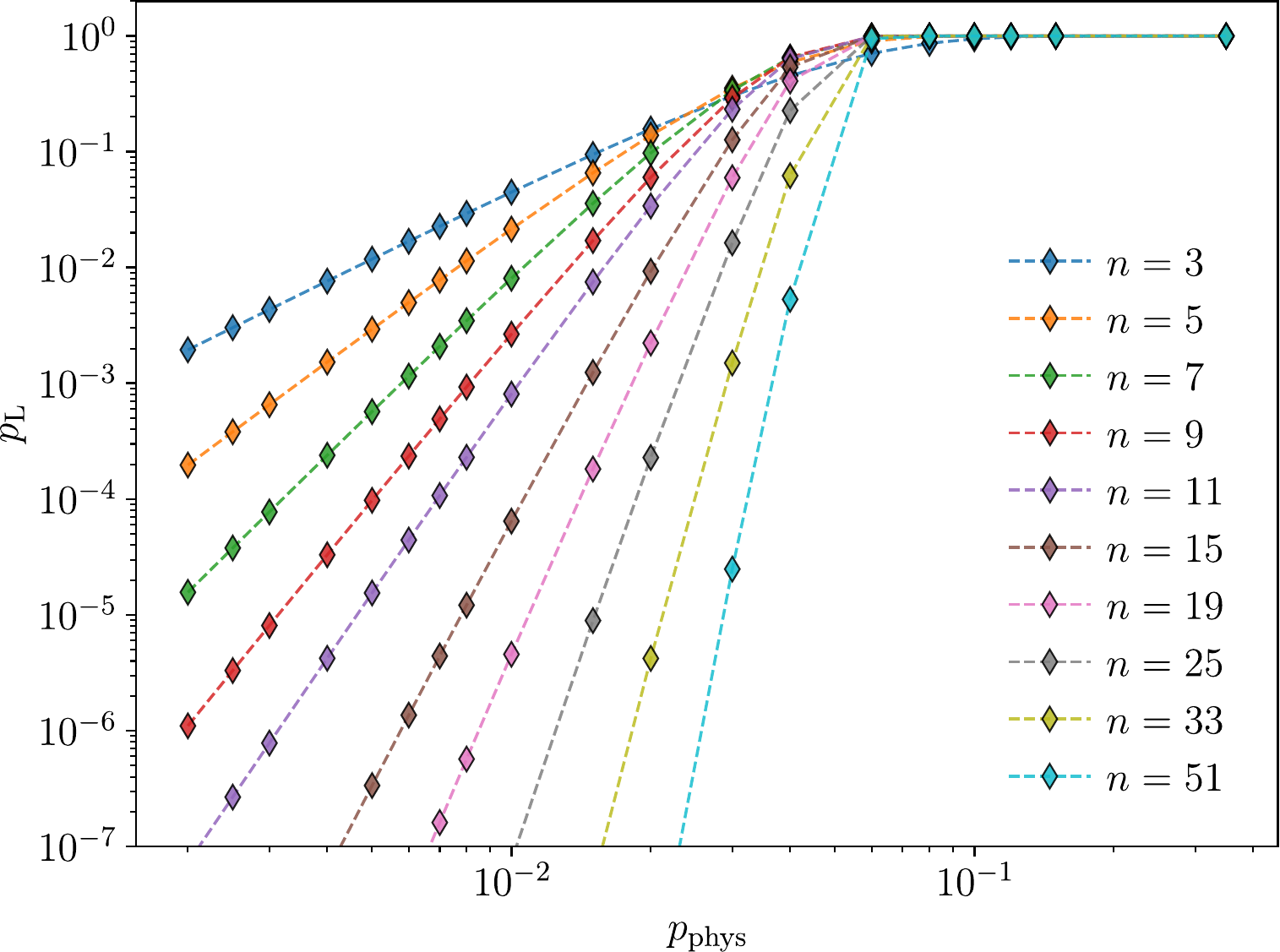}
    \caption{Double-logarithmic plot of the total logical error rate as a function of physical error rate for various $n$. The parameter $p_{\rm phys}$ describes the physical error rate per qubit during any gate, measurement or initialization and $p_L$ is the total probability for any logical errors occurring on the chip during one error correction cycle. Note that $n$ describes not only the number of logical qubits but also determines the code distance.}
    \label{fig:pl_vs_pphys}
\end{figure}
Fig.~\ref{fig:pl_vs_pphys} shows the scaling of the logical error rate with the physical error rate assuming an equal error rate $p_\text{phys}$ for all operations, i.e. ${p_1=p_2=p_\text{M}=p_\text{I}=p_\text{phys}}$, for different system sizes $n$.
The logical error rate $p_\text{L}$ increases polynomially with $p_\text{phys}$ and decays exponentially with the code distance ${d=n}$.
The results are in qualitative agreement with the results for bit-flip errors in the surface code presented in Ref.~\cite{Fowler2012}, where the same scaling is found empirically.\\

\paragraph{Non-diagonal gates}
Logical gates which are not diagonal in the $z$-basis are not fully protected. They require two CNOT chains along a logical line, meeting at a center qubit where a physical rotation is applied and propagating outwards again (see the $R_x$ gate in Fig.~1 in the main text).
These gates can introduce highly correlated errors. 
Any bit-flip errors introduced before the application of the first CNOT chain commute through the whole logical gate as a single physical error. An error occurring during or between the CNOT chains creates a chain of bit-flips starting from the affected qubit outwards to the qubit at the origin of the CNOT chain.
Given the knowledge of which logical line has been implemented before the syndrome measurement, one can thus correct such error chains as well. As these errors never affect the center qubit unless originating exactly at that qubit, all qubits along the chain can be corrected with respect to that qubit. Therefore the total logical error rate depends only on the error of the center qubit rotation and the adjacent CNOT gates, independent of the total length of the logical line.

%